\documentclass[sigconf,screen]{acmart}

\usepackage{todonotes}
\let\xtodo\todo
\renewcommand{\todo}[1]{\xtodo[inline,color=green!40]{#1}}

\usepackage{subcaption}
\usepackage{float}
\usepackage{graphicx}
\usepackage{tikz}
\usepackage{wrapfig}
\usepackage{glossaries}
\usepackage{enumitem}
\usepackage{xr}
\makeatletter
\newcommand{\setword}[2]{#1\def\@currentlabel{\unexpanded{#1}}\label{#2}}
\makeatother

\newcommand{\confabbr}{CHI '22}
\newcommand{\conffull}{\confabbr: ACM CHI Conference on Human Factors in Computing Systems}
\newcommand{\confdate}{April 30 -- May 6 2022}
\newcommand{\confloc}{New Orleans, LA, USA}

\newcommand{\maintitle}{Generative 3D Animation Pipelines}
\newcommand{\fulltitle}{\maintitle: \\Automating Facial Retargeting Workflows}

\newcommand{\authorabbr}{Girbig et al.}
\newcommand{\authorlist}{
\author{Julius Girbig}
\email{j.girbig@campus.lmu.de}
\orcid{0000-0002-8462-7390}
\affiliation{
  \institution{LMU Munich}
  \country{Germany}
}

\author{Changkun Ou}
\email{changkun.ou@ifi.lmu.de}
\orcid{0000-0002-4595-7485}
\affiliation{
  \institution{LMU Munich}
  \country{Germany}
}

\author{Sylvia Rothe}
\email{sylvia.rothe@ifi.lmu.de}
\orcid{0000-0002-3819-3608}
\affiliation{
  \institution{LMU Munich}
  \country{Germany}
}

}

\acmConference[\confabbr]{\conffull}{\confdate}{\confloc}
\acmBooktitle{\conffull}

\begin{document}
% Regarding how a position paper look like, here are many examples: \url{https://sites.google.com/view/rl4hci/position-papers}

\title[\maintitle]{\fulltitle}
\authorlist\renewcommand{\shortauthors}{\authorabbr}

\begin{teaserfigure}
    \centering
    \includegraphics[width=\textwidth]{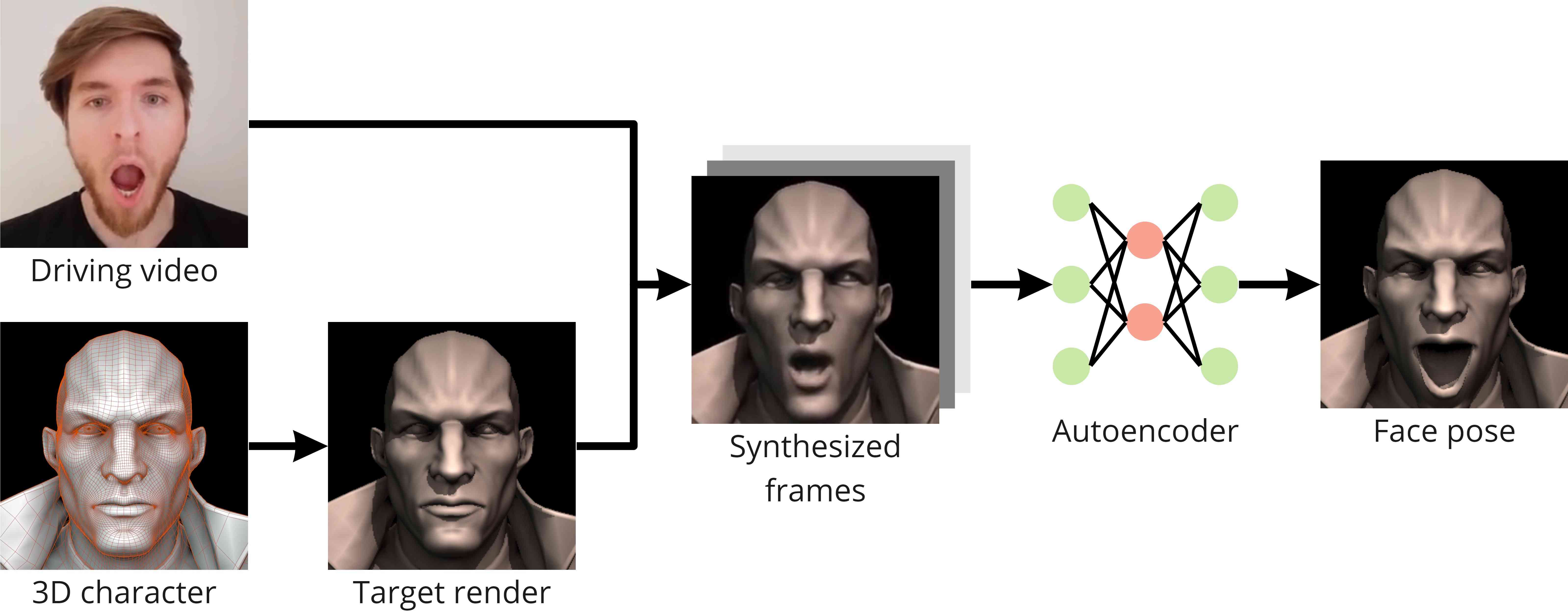}
    \caption{
    An overview of the proposed animation pipeline, which uses an RGB video of an actor and a 3D character to animate the controls of virtually any facial rig.}
    \label{fig:AlgorithmOverview}
\end{teaserfigure}

\begin{abstract}
% What is the real-world problem?
Design tools in the 3D industry, while powerful, are still tedious and labor-intensive when it comes to bringing a creative idea for a visual effect to life.
% What did we do?
In this position paper, we discussed how an infamous generative synthetic media, deepfakes, could be of use and embedded into common sophisticated 3D workflows to reduce user workloads in areas such as 3D model editing, material design, and character animation. As a case discussion, we also prototyped a tool to address the retargeting problem in character animation. 
% What did we find?
Although deepfakes themselves have received a negative public image, the results of our interviews with field experts are unexpectedly positive in regard to our tool that utilizes deepfake algorithms.
% What are the insights and implications?
Lastly, we also discussed our experience and observed design practices to put deepfakes to good use, including how we could avoid potential misuses directly by design, how this design changes user interactions, and subsequent open issues.
\end{abstract}

% http://dl.acm.org/ccs.cfm


\maketitle
\section{Introduction}

Creative arts often consist of two aspects: creation and realization.
When expressing and creating content in 3D, such as designing a 3D virtual avatar, designers and artists have to conceptualize not only a unique character but also need tedious tweaking to iterate their ideas toward perfection. 

In this process, the creation segment requires a large amount of exploration in terms of concerning dimensions, and realization exploits the design details in a design space determined by those dimensions.

In recent years, while developments in artificial intelligence (AI) have led to notorious abuses of deepfakes \cite{westerlund2019deepfake}, with increasing introspection, more and more generative machine learning models have been widely applied to the design domain to help users explore the possibilities of unseen design space, such as GANSlider \cite{dang2022ganslider}.

As a case study to explore the design principles for user elicitation, we selected a specific problem in character animation, namely facial retargeting. The situation requires an artist to tweak a pleasing facial expression to match a reference facial expression from an actual human. We developed a prototype tool that automates this tedious manual process for 3D artists to tackle this problem: First, we use an actual facial motion as a reference, then generate an altered 2D image from a rendered face image in the rest pose. With the expressive power of the deepfake, we further process the output of the deepfake using an additionally trained neural network to find a suitable transformation context that can render the transformed facial model as close to the deepfake altered picture as possible, as shown in \autoref{fig:AlgorithmOverview}.

To first evaluate our design, we discussed with practitioners the use cases for this prototype and their general impression regarding the use of deepfakes in such tools.
Surprisingly, they generally expressed themselves quite positively, inspiring us to explore other issues in the 3D field that are similar to creative arts design, such as model editing and material design.
In model editing, generative models such as deepfakes can generate samples of various potential characteristics of a model; in material design, these approaches could also create a list of candidates that inspire users and provide them with better information for selecting their preference.

For the previous three scenarios, we discuss that different and specific design principles apply to each. For example, to prevent misuse of actual facial data (or to give the impression of abuse), we might only allow non-real scenarios or object modification, and only allow identifiable outputs from such a tool.

In addition to these specific uses of deepfakes in creative art scenarios, we also argue that we should have a certain degree of concern regarding the application of these techniques and the new challenges their design brings.
\section{Case Exploration: Exploring Facial Retargeting using Deepfakes}

We approach the problem of the current 3D facial animation workflow containing large amounts of tedious and time-consuming manual labor for the animator. Challenges, such as rig incompatibility, motion data mapping, and low-resolution data can significantly increase production times. We explored the capabilities of generative learning-based algorithms by utilizing them to disconnect the facial expressions of an actor from their respective identity and transfer them to 3D characters. This process creates a 2D projection, representing a version of the desired animation based on the actor's facial expressions. The synthesized video data can be used as a reference during the animation process or to train other learning-based algorithms to animate a facial rig without the need for a user.

In order to predict positional and rotational values of animation controllers (also: bones) that drive 3D faces, we developed a prototype of a tool whose input data is composed of the 3D character and the previously mentioned synthesized video, representing a 2D version of the target animation. Our tool can also automatically register various facial rigs, making it independent from 3D characters.

With multiple tests regarding facial expressions (as seen in \autoref{fig:AlgorithmOverview}), we interviewed 10 professional 3D animators to gather insights, for instance, whether such a tool could potentially reduce animation production times. The average of the individual discussions suggest, that more than half of the time spent in 3D facial animation would be saved by using the prototype. Based on the unanimous opinion of the experts, the proposed algorithm could largely improve the current animation workflow.

The usage of a generative algorithm, namely deepfake algorithm, is crucial to the design concept and is a requirement for the method to function. While the deepfake algorithm is not implemented into the prototype, an embedded algorithm could reduce the risk of using this part of the process for malicious intents.

Even though the prototype design was subjected to multiple iterations and redesigns, there are multiple open design questions, that remain to be explored.

\begin{itemize}
    \item Could a face-swap method be used to add more landmarks, or a specific landmark layout to the face of an actor, to improve landmark-based facial tracking approaches?
    \item Could the method be extrapolated onto a full-body motion tracking solution?
    \item How much would the performance improved if the sample size of the training algorithm would be greatly increased?
    \item Is there a way to use deepfake algorithms to transform facial expressions from videos into depth maps?
\end{itemize}

\section{Additional Design Examples}
Apart from the previously described tool, there are more use cases for deepfake algorithms and other types of generative models within the 3D content creation domain. Two additional example cases are discussed in the following.

\subsection{Generative Model Editing}
Images of real-world subjects are generally 2D projections of the 3D space. We as humans can extrapolate from these images and guess the three-dimensional geometry of the scene.
\citet{nie2020total} showed that generative algorithms are capable of solving varying precision tasks by reconstructing a room with furniture in 3D space using a single image. 
Though, the capabilities of deep-learning algorithms also extend to the facial regions. SIDER has proven that even fine details of facial geometry can be reconstructed from a single input image \cite{chatziagapi2021sider}.
\citet{tran2019towards} also proposed an algorithm capable of extracting 3D face models, as well as albedo textures to generate realistic faces from single input images.

Since the previously described methods can work on single input images, we propose the design suggestion of using deepfakes and similar generative models to generate multiple face models from a single image to create a 3D expression dataset. These meshes could then be used for facial animation purposes using methods such as optical flow, to leverage realistic facial motion tracking based on scarce data.

\subsection{Generative Material Design}
As previously mentioned, \citet{tran2019towards} proposed a method that is capable of extracting albedo face textures of in-the-wild images, \citet{yang2021self} further reconstruct faces using a 3D morphable mesh and an albedo texture generator.

Generative models, such as deepfake algorithms could be used to generate missing data to improve the dataset of texture generation techniques.

An additional use case could be to generate unseen, realistic 3D characters based on an interface aimed to offer customization capabilities to artists, as shown by \citet{dang2022ganslider}. The variables set by the artist could influence the facial hair length, skin tone and other facial features.

\section{Discussion}
Using deepfake algorithms in the 3D domain can also impose several problems to the security and integrity of personal data. In the following, several design challenges and proposed methods to combat the misuse of deepfake algorithms (in our context) are proposed.

\subsection{Ways of Minimizing Potential Misuses}

To combat the possible exploitation of generative models for malicious intents, we distinguish between their direct and indirect use. In direct use, the user is allowed and required to manipulate the video data by creating the synthesized video data themselves. In the case of indirect use, the deepfake algorithm would be integrated into the program's pipeline, not directly accessible for the average user.

Moreover, a filter could be used to check the input data for malicious intent. In the case of our prototype, since the face driven by the actor's video data is a 3D model, we believe that a filter capable of differentiating between renderings and real images is feasible.

To further minimize the potential for misuse, a different data type could be generated by design. It might be beneficial to explore the use of heightmaps or vertex offset lists instead of realistic renderings, as this does not create synthesized video data at any step of the program.

\subsection{Potential Guidelines for Embedded Deepfake Algorithms}
In our case, the deepfake algorithm is driving an image of a 3D character. To reduce the risk of potential misuse, e.g. driving the face of a real human, one might modify the proposed tool to prevent the usage of real photos.

Our first suggestion is to integrate a specific generative model, that has been trained to return an invalid output if the target image of the model is a real human. The model's output could be a black image or contain a specific error message to the user.

We also propose using an additional filtering algorithm, able to additionally recognize the misuse of 3D characters. The filter could trigger a halt within the processing and force the program to exit immediately. While there is a point to be made about this method being safer than the previously mentioned guideline, the separation of the deepfake and the filter might allow users to disconnect the filter from the program and bypass this security measure.

Watermarking content to prevent the use of deepfake algorithm for malicious intents, is not a novelty. \citet{alattar2020a} proposed a system where trusted news entities embed unique digital watermarks in their videos, to facilitate a method to retrieve the watermark from the content and match the watermark to the original video. A similar system could be utilized in our case, by watermarking the produced synthesized video content and recording the history and provenance of the input content in a blockchain, as described by Alattar et al.

Finally, the option of refraining from implementing a deepfake algorithm into the tool still remains. While this would require a modular system, where the generative model can be exchanged easily, the provider of the deepfake plugin retains full control over the kind of software they distribute. This could enable system, where every user needs to authenticate themselves before they are allowed to download the algorithm plugin. We presume that the performance of the forensic analysis of malicious use cases will increase, as a result.

\subsection{Collaboration between Users and Tools}

Inspired by the aforementioned omission of the implementation of a deepfake algorithm, we argue that a modular tool structure could potentially be beneficial for users. As the quality and capabilities of deepfake algorithms are advancing rapidly, users would be able to exchange the included algorithm with a more recent version. This could not only increase the lifetime of the program, but also grant access to new features.

During the interviews with the professionals, controls for computational customizations were considered. For instance, a slider to scale the intensity of the output animations of our developed prototype could be used, as well as controls for the generative power of the deepfake model could greatly improve the usability and output quality of the tool.

Another form of customization could be artistic modifications to the model parameters. \citet{dang2022ganslider} researched the implications of different slider setups, where the styling of an image of a human face could be adjusted. This could open up artistic freedom, as in our proposed tool, a 3D animator would be able to adjust the type of facial expression using a slider. This could enable fine control over the animation output, such as adjusting the corners of a smiling mouth or switching the animation style from realistic to cartoon-like.

One of the more involved collaboration techniques is the feedback loop. In our case, the prototype would predict the facial pose based on an input image, though the animator would be able to adjust the pose to their personal preferences. The algorithm would include the changes made by the human in their next processing loop and create new poses considering the artist's feedback.

\subsection{Remaining Abuses and Challenges}
While we proposed multiple methods to combat the intentional misuse of deepfake algorithms, some problems remain and new possibilities to exploit the proposed tool arise.

Since the prototype is able to work with 3D animated characters, licensed content such as trademarked characters could be used without permission. Users would be able to produce media content of the 3D characters, provided they gain access to the rigged meshes themselves.

A more sensitive and severe problem is the distortion of 3D characters and their actions into inappropriate scenarios. This issue could range from creating horror movies based on characters aimed at children to intentionally scare them, to users creating distorted content such as pornography.

This shows that it is still possible to generate malicious media content, although we argue that our developed prototype stands as a positive example of a good use case for deepfake algorithms and generative models overall. We leave these open issues to further ethics discussions and legal developments.

\begin{acks}
The authors of this paper would like to thank Karo Castello for the insightful discussion, which provided an application-oriented perspective to the initial concept; Prof. Andreas Butz for supporting this research.
\end{acks}

\bibliographystyle{ACM-Reference-Format}
\bibliography{ref}

\end{document}